\documentclass{jnmp}

\usepackage{amsmath}

\JNMPnumberwithin{equation}{section}

\allowdisplaybreaks

\theoremstyle{definition}

\begin{document}

\resetfootnoterule

\renewcommand{\evenhead}{PGL Leach and J Miritzis}
\renewcommand{\oddhead}{Analytic Behaviour of Competition among Three Species}

\thispagestyle{empty}

\FirstPageHead{*}{*}{2006}{\pageref{firstpage}--\pageref{lastpage}}{Article}

\copyrightnote{2006}{PGL Leach and J Miritzis}

\Name{Analytic Behaviour of Competition among Three Species}

\label{firstpage}

\Author{PGL Leach~$^\dag$\footnote{permanent address: School of
Mathematical Sciences, Howard College, University of
KwaZulu-Natal, Durban 4041, Republic of South Africa} and J
Miritzis}

\Address{Department of Marine Sciences, University of the Aegean\\Mytilene 81 100, Greece\\Email: imyr@aegean.gr; leachp@ukzn.ac.za}

\Date{Received , 2005; Revised Month *, 200*;
Accepted Month *, 200*}

\begin{abstract}
We analyse the classical model of competition between three species studied by
May and Leonard ({\it SIAM J Appl Math} \textbf{29} (1975) 243-256) with the
approaches of singularity analysis and symmetry analysis to identify values of
the parameters for which the system is integrable. We observe some striking
relations between critical values arising from the approach of dynamical
systems and the singularity and symmetry analyses.

\end{abstract}

\section{Introduction}

In a classic study of a model of competition among three species May and Leonard \cite{may75a} demonstrated the dramatic change in the qualitative
behaviour of the model in simply going from two species to three species. The
Gausse-Lotka-Volterra \cite{gausse34a,lotka25a,volterra26a} model for competition among $n$ species is
\begin{equation}
\dot{N}_{i}=r_{i}N_{i}\left(  1- {\displaystyle\sum\limits_{j=1}^{n}} a_{ij}N_{j}\right),\ \ \ \ i=1,...,n, \label{l-v}
\end{equation}
where $N_{i}\left(  t\right)  $ is the size of population $i$ at time $t,$
$r_{i}$ is its intrinsic growth rate, $a_{ij}$ the coefficient representing
the effect on its growth rate due to species $j$ and overdot denotes
differentiation with respect to time. May and Leonard restrict the number of
competing species to three and make some assumptions about the parameters in
the system to reduce the system to one which is susceptible to analytic
treatment in the main. The critical point in choosing $n=3$ is not the
smallness of the number, but the potential for dramatic change in the
behaviour of the system in going from $n=2$ to $n=3.$ When $n=2,$ the
autonomous system, (\ref{l-v}) with $n=2,$ can be reduced to a single
first-order equation and is integrable. This is not automatically the case for
$n=3.$ Indeed the potential, if not its realisation, for chaos exists. The
parameters, $r_{i},\,\,i =1,n $, are taken to be equal and then set at unity
by a rescaling of time. Rescaling of the independent variables enables the
diagonal elements of the quadratic terms to be set at unity. Finally the
interaction coefficients are limited to just two by the assumptions that
species $i+1(\operatorname{mod}3)$ affects species $i,$ $i=1,2,3,$ and species
$i+2(\operatorname{mod}3)$ affects species $i,$ $i=1,2,3,$ in the same way.
The model system is then
\begin{align}
\dot{x}  &  =x\left(  1-x-\alpha y-\beta z\right),\nonumber\\
\dot{y}  &  =y\left(  1-\beta x-y-\alpha z\right),\label{3-d}\\
\dot{z}  &  =z\left(  1-\alpha x-\beta y-z\right).\nonumber
\end{align}
We remark that the simplifications made to the values of the parameters are
not as restrictive as one would imagine. For example in grasslands the
reproductive rates of different species of ungulants of similar size are
expected to be similar and the coefficients of competition likewise. Indeed
under good grazing conditions the $a_{ij}$ would be anticipated to be low and
of comparable magnitude. The mathematical attraction of this model is that the
community matrix, \textit{videlicet}
\begin{equation}
A=\left[
\begin{array}
[c]{lll}%
1 & \alpha & \beta\\
\beta & 1 & \alpha\\
\alpha & \beta & 1
\end{array}
\right], \label{commu}%
\end{equation}
is a circulant matrix for which an explicit formula for the eigenvalues exists
\cite{berlin52a}. With the entries of $A$ as indicated its eigenvalues are
\begin{align*}
\lambda_{1}  &  =1+\alpha+\beta,\\
\lambda_{2\pm}  &  =\frac{1}{2}\left[  2-\alpha-\beta\pm i\sqrt{3}\left(
\alpha-\beta\right)  \right].
\end{align*}

May and Leonard give the equilibrium points of system (\ref{3-d}) as $\left(
0,0,0\right)$; $\left(  1,0,0\right)$, $\left(  0,1,0\right)  $ and $\left(
0,0,1\right)$; $\left(  1-\alpha,1-\beta,0\right)  /\gamma$, $\left(
1-\beta,0,1-\alpha\right)  /\gamma$ and $\left(  0,1-\alpha,1-\beta\right)
/\gamma$, where $\gamma= 1 - \alpha\beta$; $\left(  1,1,1\right)  /\left(
1+\alpha+\beta\right)  $ for zero-, one-, two- and three-population
equilibria. In this paper we investigate the properties of system (\ref{3-d})
from the approaches of singularity analysis and symmetry analysis. We
emphasise that the thrust of our investigations is the integrability of system
(\ref{3-d}) and not its qualitative behaviour for which the methods of
dynamical systems are well-suited. The singularity analysis is directed
towards the determination of the existence of solutions which are analytic.
Symmetry analysis leads towards invariance of the system under infinitesimal
transformation so that in the presence of a suitable number of symmetries the
solution of the system may be reduced to a sequence of quadratures or the
existence of three functionally independent invariants from which the solution
follows by a process of elimination of variables. In the case of the latter
the elimination may be only local through the use of the Implicit Function
Theorem. Equally the performance of the quadratures may not be possible in
closed form or lead to analytic solutions.

Before we begin any analysis we observe that under a constraint upon the
parameters $\alpha$ and $\beta$ system (\ref{3-d}) is an example of a
decomposed system since, if we add the three equations, we have
\begin{equation}
(x+y+z) ^{\mathbf{.}} = (x+y+z) -\left\{  x ^{2}+y ^{2}+z ^{2}+ (\alpha+\beta)
(xy +yz +zx)\right\}. \label{4.111}%
\end{equation}
Clearly the constraint $\alpha+\beta= 2 $ enables us to write (\ref{4.111}) as
the composed system
\begin{equation}
\dot{u}=u-u^{2}, \label{ricat}%
\end{equation}
where $x+y+z = u $, which is readily integrated to give the invariant
\begin{equation}
I_{1}=\left[  \frac{1}{x+y+z}-1 \right]  e^{t} \label{inva}%
\end{equation}
and this can be rearranged as the analytic solution%
\begin{equation}
x+y+z=\frac{\mbox{\rm e} ^{t}} {I_{1}+ \mbox{\rm e} ^{t}} \label{sum}%
\end{equation}
for the total population.
(Recall that time was rescaled; this explains the simplicity of the time
dependence in (\ref{sum})). One of the attractive features of decomposible
systems is that the composed equation, particularly in the case of systems of
first-order differential equation of the type usually encountered in
modelling, is usually integrable so that an invariant exists and the dimension
of the system is effectively reduced by one
\cite{andriopoulos02a,andriopoulos03a,andriopoulos05a}.

\section{Singularity analysis of system (\ref{3-d})}

We follow the standard method of singularity analysis\footnote{The reader is
referred to Ramani \textit{et al} \cite{ramani89a} and Tabor \cite{tabor89a}
for an account of the details of the application of the Painlev\'e Test and
implementation of the ARS algorithm.} and determine the leading-order
behaviour by setting $x=A\tau^{p},$ $y=B\tau^{q}$ and $z=C\tau^{r},$ where
$\tau=t-t_{0}$ and $t_{0}$ is the location of the putative movable pole, in
system (\ref{3-d}) to obtain%
\begin{align*}
pA\tau^{p-1}  &  =A\tau^{p}\left(  1-A\tau^{p}-\alpha B\tau^{q}-\beta
C\tau^{r}\right),\\
qB\tau^{q-1}  &  =B\tau^{q}\left(  1-\beta A\tau^{p}-B\tau^{q}-\alpha
C\tau^{r}\right),\\
rC\tau^{r-1}  &  =C\tau^{r}\left(  1-\alpha A\tau^{p}-\beta B\tau^{q}%
-C\tau^{r}\right),
\end{align*}
from which it is evident that the linear terms of the right hand side are not
to be considered for the determination of the leading-order behaviour or of
the resonances. On the assumption that the leading-order behaviour assumed
does in fact represent polelike behaviour in the three dependent variables the
requirement of balance of the terms reduces to just $-1,p,q,r$ from which it
is evident that $p=q=r=-1.$ The possibility that one or other of the exponents
differs from $-1$ cannot be entertained since the exponent would then be
nonnegative and the singularity analysis does not admit such a possibility for
integral leading-order behaviour. One could imagine the introduction of branch
point singularities with fractional exponents, but this leads us away from the
standard analysis.

With the common exponent of the leading-order behaviour being $-1$ the
coefficients of the leading-order terms are found from the solution of the
system
\[
\left[
\begin{array}
[c]{lll}%
1 & \alpha & \beta\\
\beta & 1 & \alpha\\
\alpha & \beta & 1
\end{array}
\right]  \left[
\begin{array}
[c]{l}%
A\\
B\\
C
\end{array}
\right]  =\left[
\begin{array}
[c]{l}%
1\\
1\\
1
\end{array}
\right]  \Rightarrow\left[
\begin{array}
[c]{l}%
A\\
B\\
C
\end{array}
\right]  =\frac{1}{1+\alpha+\beta}\left[
\begin{array}
[c]{l}%
1\\
1\\
1
\end{array}
\right]
\]
and we note that the solution is the same as the location of the interior
equilibrium point of the system (\ref{3-d}).

The resonances are determined from system (\ref{3-d}) (with the linear term
omitted) by the substitutions
\[
x=A\tau^{-1}+M\tau^{r-1},\ \ y=B\tau^{-1}+N\tau^{r-1},\ \ z=C\tau^{-1}%
+S\tau^{r-1},
\]
where $r$ denotes, as usual, the resonance and should not been confused with
the usage above as one of the exponents of the leading-order behaviour. The
terms linear in $M $, $N$ and $S$ give the eigenvalue problem%
\begin{equation}
\left[
\begin{array}
[c]{lll}%
r+A & \alpha A & \beta A\\
\beta B & r+B & \alpha B\\
\alpha C & \beta C & r+C
\end{array}
\right]  \left[
\begin{array}
[c]{l}%
M\\
N\\
S
\end{array}
\right]  =\left[
\begin{array}
[c]{l}%
0\\
0\\
0
\end{array}
\right]  \label{eige}%
\end{equation}
from which it follows that
\begin{equation}
r_{1}=-1\ \ \mathrm{and}\ r_{2\pm}=\frac{1}{2}\left[  \frac{\alpha+\beta-2\pm
i\sqrt{3}\left(  \alpha-\beta\right)  }{1+\alpha+\beta}\right]. \label{reso}%
\end{equation}

If we look at the eigenvalues of the Jacobian matrix of system (\ref{3-d}) at
the interior equilibrium, we find that
\[
\lambda_{i}=-\left(  1+\alpha+\beta\right)  r_{i},\ \ i=1,2\pm.
\]
May and Leonard have a stable equilibrium point with three species present if
$\alpha+\beta<2$ ($\alpha,\beta>0$). For the region $\alpha>1$ and $\beta>1$
the equilibrium points for all three single species are each stable. For the
remaining points $\left(  \alpha,\beta\right)  $ of the parametric space
asymptotically stable equilibrium points do not exist.

In (\ref{reso}) the singularity analysis demonstrates that there is no
possibility of an analytic solution unless $\alpha=\beta$ for otherwise
$r_{2\pm}$ are complex. When $\alpha=\beta$, $r_{2\pm}$ coalesce into%
\[
r_{2}=\frac{\alpha-1}{1+2\alpha},
\]
with $r_{2}$ a positive integer, $n$, if
\[
\alpha=-\frac{n+1}{2n-1},
\]
which is necessarily negative and so beyond the acceptable parameter range of
the model. Only in the case that $\alpha\left(  =\beta\right)  =1$, for which
$r_{2}=0\left(  2\right)  $ can we expect an analytic solution. Then
(\ref{eige}) is
\begin{equation}
\left[
\begin{array}
[c]{lll}%
1 & 1 & 1\\
1 & 1 & 1\\
1 & 1 & 1
\end{array}
\right]  \left[
\begin{array}
[c]{l}%
M\\
N\\
S
\end{array}
\right]  =\left[
\begin{array}
[c]{l}%
0\\
0\\
0
\end{array}
\right]  \Rightarrow\left[
\begin{array}
[c]{l}%
M\\
N\\
S
\end{array}
\right]  =k_{1}\left[
\begin{array}
[c]{r}%
1\\
0\\
-1
\end{array}
\right]  +k_{2}\left[
\begin{array}
[c]{r}%
0\\
1\\
-1
\end{array}
\right], \label{mns}%
\end{equation}
where $k_{1}$ and $k_{2}$ are arbitrary parameters, and we have the two
constants of integration entering at the leading-order behaviour. The linear
terms omitted in the analysis of the dominant terms do not cause an
inconsistency since they do not affect the leading order term of $\tau^{-2}.$

That the solution is analytic may be demonstrated by the explicit integration
of system (\ref{3-d}) with $\alpha=\beta=1.$ When the composed system is
integrated and this is substituted into the equations for $x$ and $y,$ say,
\textit{ie} $z$ is eliminated using the invariant, the equations for $x$ and
$y$ decouple and one obtains the solution by simple quadratures to be%
\[
x=\frac{k_{1}e^{t}}{C+e^{t}},\ \ y=\frac{k_{2}e^{t}}{C+e^{t}},\ \ z=\frac
{1+\left(  1-k_{1}-k_{2}\right)  e^{t}}{C+e^{t}},
\]
as is suggested by the solutions given in (\ref{mns}).

In terms of the singularity analysis system (\ref{3-d}) is
integrable in terms of analytic functions at the specific point
$\left(  1,1\right)  $ in the $\left(  \alpha,\beta\right)  $
plane. This is the point of contact between the two regions of
stable equilibria reported by May and Leonard \cite{may75a} [Fig
1]. We observe that the only pattern of leading-order behaviour
compatible with the standard method of singularity analysis as
found in, say, \cite{ramani89a,tabor89a} is that the exponents of
the leading-order terms be at $-1.$ However, we may
depart\footnote {We do not claim any originality in making a
departure.  Daniel {\it et al} \cite {Daniel 92 a} did the same in
their study of the Heisenberg spin chain with anisotropy and
transverse field.  For a deep study from the viewpoint of
cosmological interests see the more recent work by Cotsakis \cite
{Cotsakis 06 a}.}  from that standard analysis and investigate the
consequences. If we suppose that $p=q=-1$ and $r=0,$ the dominant
terms of
system (\ref{3-d}) become%
\begin{align}
-A\tau^{-2}  &  =A\tau^{-1}\left(  -A\tau^{-1}-\alpha B\tau^{-1}\right),\nonumber\\
-B\tau^{-2}  &  =B\tau^{-1}\left(  \beta A\tau^{-1}-B\tau^{-1}\right),\label{r0}\\
0  &  =C\left(  -\alpha A\tau^{-1}-\beta B\tau^{-1}\right).\nonumber
\end{align}
The first two of (\ref{r0}) give
\[
\left[
\begin{array}
[c]{ll}%
1 & \alpha\\
\beta & 1
\end{array}
\right]  \left[
\begin{array}
[c]{l}%
A\\
B
\end{array}
\right]  =\left[
\begin{array}
[c]{l}%
1\\
1
\end{array}
\right]  \Rightarrow\left[
\begin{array}
[c]{l}%
A\\
B
\end{array}
\right]  =\frac{1}{1-\alpha\beta}\left[
\begin{array}
[c]{l}%
1-\alpha\\
1-\beta
\end{array}
\right].
\]
The third of (\ref{r0}) gives either $C=0$ or $\alpha A+\beta B=0.$ This
second condition, coupled with the first and second of (\ref{r0}) demands
either that $\alpha\beta=1\Rightarrow\alpha=1,\beta=1,$ or places $\alpha$ and
$\beta$ on the circle%
\[
\left(  \alpha-\frac{1}{2}\right)  ^{2}+\left(  \beta-\frac{1}{2}\right)
^{2}=\frac{1}{2}%
\]
in the $(\alpha,\beta) $ plane. The former condition, $C =0 $, coincides with
one of the equilibrium points with just two species, $x$ and $y$, present. The
other two possibilities, \textit{videlicet} $p=0,$ $q=r=-1$ and $p=-1,$ $q=0$
and $r=-1$, correspond to the equilibrium points with two solutions given by
$\left(  0,1-\alpha,1-\beta\right)  /\left(  1-\alpha\beta\right)  $ and
$\left(  1-\beta,1-\alpha,0\right)  /\left(  1-\alpha\beta\right)  $ respectively.

In a similar situation, if we take the leading-order exponents to be $p=-1,$
$q=r=0,$ we obtain the leading-order behaviour,%
\begin{align*}
-A\tau^{-2}  &  =A\tau^{-1}\left(  -A\tau^{-1}\right),\\
0  &  =B\left(  \beta A\tau^{-1}\right),\\
0  &  =C\left(  -\alpha A\tau^{-1}\right),
\end{align*}
for which the solution is obviously $\left(  1,0,0\right)$.   The other two
possibilities are $\left(  0,1,0\right)  $ and $\left(  0,0,1\right)  ,$
\textit{ie} we recover the equilibrium points with just a single species present.

Moreover, if one makes a formal expansion
\[
x =\sum_{i = 0} ^ {\infty} a_i\tau ^ {i- 1},\quad y =\sum_{i = 0} ^ {\infty} b_i\tau ^ {i- 1},\quad z =\sum_{i = 0} ^ {\infty} c_i\tau ^ {i- 1}
\]
and substitutes this into (\ref {3-d}), one obtains (courtesy of Mathematica) that $b_i =0 $, $c_i =0 $, $\forall i $, all odd coefficients, $a_{2i+ 1} $, are zero and that
\[
a_0 =1,\,\, a_1 = \frac {1} {2},\,\, a_2=\frac {1} {12},\,\, a_4 = -\frac {1} {720},\,\, a_6 = \frac {1} {30240},\,\, a_8 = -\frac {1} {1209600},\,\, a_{10} = \frac {1} {47900160}
\]
which is in accordance with the solution of one species present, {\it videlicet}
\[
x = \frac {1} {1 -\exp [ t-t_0]},\quad y =0,\quad z =0.
\]
Here we must emphasise again that we are not applying singularity analysis in the sense of the Painlev\'e Taste.  Nevertheless we see an interesting connection between the results of dynamical systems analysis and the simple series substitution. The solution obtained is consistent with the Painlev\'e analysis in that it leads to a subsidiary solution \cite {Rajasekar 04 a} although the route to its obtention is formally different.  Nevertheless it cannot be regarded as a subset of the Painlev\'e analysis since a fundamental feature of the analysis is that the coefficients of the leading order terms be nonzero.

We emphasise that this last part of the analysis is not in accordance with the
norms of singularity analysis as presented in the standard references. Once we
admit the possibility of a zero as the exponent of the leading-order behaviour
of one or more species, we depart from the criteria for the application of the
Painlev\'{e} test. Nevertheless we see that results can be obtained which are
very suggestive and which connect in a natural way with the analysis of system
(\ref{3-d}) via dynamical systems. For an investigation of the presence of two
competing species we refer the reader to \cite{lemi04}.

We conclude our singularity analysis of system (\ref{3-d}) with the final
observation that generically (\ref{3-d}) is not integrable in terms of
analytic functions. Nevertheless the analysis has revealed aspects of the
properties of the system which perhaps would not be anticipated a priori.

\section{Symmetry analysis}

The system (\ref{3-d}) is autonomous and so possesses the Lie point symmetry
$\partial_{t}.$ For integrability in the sense of Lie we require the knowledge
of a three-dimensional solvable algebra. The knowledge that (\ref{3-d}) is a
system of first-order differential equations and so possesses an infinite
number of Lie point symmetries does not help us to find the additional two
symmetries. For the purposes of symmetry analysis we make a change of
variables to convert system (\ref{3-d}) to a quadratic system\footnote{For a
system of the general form of (\ref{l-v}) this is not possible, but the
simplifying assumptions of May and Leonard that $r_{i}=r\rightarrow1,\,\,i =
1,n, $ under a rescaling of time does enable the transformation of system
(\ref{3-d}) to the simpler form. The change of variables used here is not
beneficial for the singularity analysis the outcome of which is very much
dependent upon the representation of the coordinates used, but, as it is a
point transformation, has no effect upon the algebraic structure of the
system.}.

We write
\[
X=x\mbox{\rm e}^{-t},\ Y=y\mbox{\rm e}^{-t},\ Z=z\mbox{\rm e}^{-t},\ T=\mbox{\rm e}^{t}.
\]
Then system (\ref{3-d}) becomes%
\begin{align}
X^{\prime}  &  =-X\left(  X+\alpha Y+\beta Z\right),\nonumber\\
Y^{\prime}  &  =-Y\left(  \beta X+Y+\alpha Z\right),\label{quad}\\
Z^{\prime}  &  =-Z\left(  \alpha X+\beta Y+Z\right),\nonumber
\end{align}
where we use the prime to denote differentiation with respect to the `new
time', $T.$   If one assumes a solution of the form $X\propto T^{p}$, $Y\propto T^{q}$ and $Z\propto
T^{r}$, one finds that
\begin{equation}
\left[
\begin{array}
[c]{c}%
X\left(  T\right) \\
Y\left(  T\right) \\
Z\left(  T\right)
\end{array}
\right]  =\frac{T^{-1}}{1+\alpha+\beta}\left[
\begin{array}
[c]{c}%
1\\
1\\
1
\end{array}
\right], \label{sol}%
\end{equation}
is a solution.  It is not the general solution because it
does not depend upon three arbitrary constants.  It is simply a peculiar
solution. In fact, going back to the original variables%
\[
X=x\mbox{\rm e}^{-t},\ Y=y\mbox{\rm e}^{-t},\ Z=z\mbox{\rm e}^{-t}%
,\ T=\mbox{\rm e}^{t},
\]
solution (\ref{sol}) corresponds to the equilibrium solution
\[
\frac{1}{1+\alpha+\beta}\left(  1,1,1\right)  ^{T}%
\]
of the original system (\ref{3-d}). Note also that to every equilibrium solution of
the original system (\ref{3-d}) there corresponds a peculiar solution of (\ref{sol}).

By inspection (\ref{quad}) possesses the two Lie point symmetries%
\begin{equation}
\Gamma_{1}=\partial_{T},\ \Gamma_{2}=-T\partial_{T}+X\partial_{X}%
+Y\partial_{Y}+Z\partial_{Z} \label{gama}%
\end{equation}
with the Lie bracket $\left[  \Gamma_{1},\Gamma_{2}\right]  _{LB}=-\Gamma_{1}.$

In analogy with (\ref{4.111}) we add the constituent equations of (\ref{quad})
to obtain
\begin{equation}
(X +Y+Z)^{\prime}= - \left\{  X ^{2}+ (\alpha+\beta)XY +Y ^{2}+ (\alpha
+\beta)YZ +Z ^{2}+ (\alpha+\beta)ZX\right\}. \label{20.1}%
\end{equation}
In the particular case that $\alpha+\beta=2 $ a possible source of additional
symmetry is from the decomposition of symmetries of the composed system of
(\ref{quad}), \textit{videlicet}%
\begin{equation}
N^{\prime}+ N^{2}=0, \label{sum1}%
\end{equation}
where $N=X+Y+Z$. We are unaware of this approach being used in the literature
before now.

As a first-order differential equation (\ref{sum1}) has the same problem of
determination of symmetries as the composed system (\ref{quad}), but it can be
written as the second-order differential equation%
\begin{equation}
w^{\prime\prime}=0, \label{w}%
\end{equation}
by means of the Riccati transformation%
\[
N= \frac{w^{\prime}}{w}\Leftrightarrow w=\exp\left(  \int N\mbox{\rm d}
T\right)  ,\ w^{\prime}= N\exp\left(  \int N\mbox{\rm d} T\right).
\]
A symmetry $\Sigma= \Theta\partial_{T}+\Xi\partial_{w}$ of (\ref{w}) can be
written as a symmetry of (\ref{sum1}), $\Lambda=\tau\partial_{T}+\eta
\partial_{N},$ as follows. The first extension of $\Sigma$%
\[
\Sigma^{\left[  1\right]  }=\Theta\partial_{T}+\Xi\partial_{w}+\left(
\Xi^{\prime}-w^{\prime}\Theta^{\prime}\right)  \partial_{w^{\prime}%
}\rightarrow\Theta\partial_{T}+\left(  \frac{d}{dT}\left(  \frac{\Xi}%
{w}\right)  -N\Theta^{\prime}\right)  \partial_{N},
\]
so that
\[
\tau=\Theta,\ \ \ \eta=\frac{d}{dT}\left(  \frac{\Xi}{w}\right)
-N\Theta^{\prime}.
\]
The Lie point symmetries of (\ref{w}) transform as follows%
\[%
\begin{array}
[c]{lll}%
\Sigma_{1}=\partial_{w} & \rightarrow & \Lambda_{1}=N\exp\left(  -\int
N\mbox{\rm d} T\right)  \partial_{N}\\
\Sigma_{2}=\tau\partial_{w} & \rightarrow & \Lambda_{2}=\left(  {1}-
{}TN\right)  \exp\left(  -\int N\mbox{\rm d} T\right)  \partial_{N}\\
\Sigma_{}=w\partial_{w} &  &
\mathrm{lost\ (The\ source\ of\ the\ Riccati\ transfor}\text{mat}%
\mathrm{ion)}\\
\Sigma_{4}=\partial_{T} & \rightarrow & \Lambda_{4}=\partial_{T}\\
\Sigma_{5}=2T\partial_{T}+w\partial_{w} & \rightarrow & \Lambda_{5}%
=T\partial_{T}-N\partial_{N}\\
\Sigma_{6}= T^{2}\partial_{T}+Tw\partial_{w} & \rightarrow & \Lambda_{6}=
T^{2}\partial_{T}+\left(  1 -2TN\right)  \partial_{N}\\
\Sigma_{7}=w\partial_{T} & \rightarrow & \Lambda_{7}=\exp\left(  \int
N\mbox{\rm d} T\right)  \left(  \partial_{T}-N^{2}\partial_{N}\right) \\
\Sigma_{8}=Tw\partial_{T}+w^{2}\partial_{w} & \rightarrow & \Lambda_{8}%
=\exp\left(  \int N\mbox{\rm d} T\right)  \left(  T\partial_{T}-TN^{2}%
\partial_{N}\right).
\end{array}
\]
In $\Lambda_{4}$ and $\Lambda_{5}$ we have the $\Gamma_{1}$ and $\Gamma_{2}$
of (\ref{gama}). If we examine the remaining symmetries, $\Lambda_{2}$ and
$\Lambda_{6}$ do not decompose. The remaining symmetries decompose according
to%
\[%
\begin{array}
[c]{lll}%
\Lambda_{1} & \rightarrow & \Delta_{1}^{\left[  1\right]  }=\exp\left(  -\int
N\mbox{\rm d} T\right)  \left\{  X\partial_{X}+Y\partial_{Y}+Z\partial
_{Z}+2\left(  X^{\prime}\partial_{X^{\prime}}+Y^{\prime}\partial_{Y^{\prime}%
}+Z^{\prime}\partial_{Z^{\prime}}\right)  \right\} \\
\Lambda_{7} & \rightarrow & \Delta_{7}^{\left[  1\right]  }=\exp\left(  \int
N\mbox{\rm d} T\right)  \left\{  \partial_{T}-N\left[  X\partial_{X}%
+Y\partial_{Y}+Z\partial_{Z}+2\left(  X^{\prime}\partial_{X^{\prime}%
}+Y^{\prime}\partial_{Y^{\prime}}+Z^{\prime}\partial_{Z^{\prime}}\right)
\right]  \right\} \\
\Lambda_{8} & \rightarrow & \Delta_{8}^{\left[  1\right]  }=\exp\left(  \int
N\mbox{\rm d} T\right)  \left\{  T\partial_{T}-TN\left[  X\partial
_{X}+Y\partial_{Y}+Z\partial_{Z}+2\left(  X^{\prime}\partial_{X^{\prime}%
}+Y^{\prime}\partial_{Y^{\prime}}+Z^{\prime}\partial_{Z^{\prime}}\right)
\right]  \right\},
\end{array}
\]
where we have written the first extensions of the decomposed symmetries to
highlight the discomforting point that these three symmetries bring no new
information. The effects of $\Delta_{1},\Delta_{7}$ and $\Delta_{8}$ are the
same as that of $\Gamma_{2}\left(  \Leftrightarrow\Lambda_{6}\right)  $ on the
autonomous system. We conclude that it is possible to decompose symmetries
just as it is possible to decompose equations, but the results are not
necessarily useful.

We already have an invariant derived from the composition of system
(\ref{3-d}) with $\alpha+\beta=2 $ in (\ref{inva}), \textit{videlicet}%
\[
I_{1}=\frac{1}{X+Y+Z}- T,
\]
when written in the new coordinates. It is evident that $I_{1}$ is not an
invariant associated with either $\Gamma_{1}$ or $\Gamma_{2}$ and so we may
use $\Gamma_{1}$ and $\Gamma_{2}$ to seek a new invariant, in fact an integral
if we require the function to vanish under the action of both
symmetries\footnote{Although one usually looks for an integral/invariant
associated with a a single symmetry -- the only way possible for a two
dimensional system -- there are at times great benefit and simplification to
imposing the requirement that the integral/invariant be associated with two
(or more) symmetries \cite{cotsakis98a}.}.

In essence we use the method of reduction of order \cite{nucci96a,nucci01a}
with the two symmetries $\Gamma_{1}$ and $\Gamma_{2}.$ The former is a
consequence of the autonomy of system (\ref{quad}) and we eliminate $T$ as the
independent variable in favour of $Z$ by writing\footnote{There is no
essential difference made by the particular choice of a new independent
variable.}%
\begin{equation}
\frac{dX}{dZ}=\frac{X\left(  X+\alpha Y+\beta Z\right)  }{Z\left(  \alpha
X+\beta Y+Z\right)  },\ \ \ \ \frac{dY}{dZ}=\frac{Y\left(  \beta X+Y+\alpha
Z\right)  }{Z\left(  \alpha X+\beta Y+Z\right)  }.\ \label{2-d}%
\end{equation}
System (\ref{2-d}) is homogeneous with the obvious symmetry $\tilde{\Gamma
}_{2}=X\partial_{X}+Y\partial_{Y}+Z\partial_{Z}$ following from $\Gamma_{2}.$
Under the standard change of variables%
\[
X=uZ,\ \ Y=vZ,\ \ \tilde{\Gamma}_{2}\rightarrow Z\partial_{Z}%
\]
we may eliminate the ignorable coordinate $Z$ (actually in the form
$\exp\left(  -Z\right)  $) to obtain the single first-order differential
equation%
\begin{equation}
\frac{dv}{du}=\frac{v\left[  \left(  \beta-\alpha\right)  u+\left(
1-\beta\right)  v+\left(  \alpha-1\right)  \right]  }{u\left[  \left(
1-\alpha\right)  u+\left(  \alpha-\beta\right)  v+\left(  \beta-1\right)
\right]  }. \label{1-d}%
\end{equation}
It is a trivial matter to integrate (\ref{1-d}) in the case $\alpha=\beta$
which, we recall, is the condition for the singularity analysis to give real
resonances. The integral is%
\[
I_{2}=\frac{v\left(  u-1\right)  }{u\left(  v-1\right)  },
\]
in which we note that the parameter $\alpha$ is absent.

When $\alpha\not =\beta$, the first-order differential equation is not
integrable in closed form (as far as Mathematica is concerned). When
$\alpha+\beta=2$, the integration of (\ref{sum1}) gives
\begin{equation}
J_{1}=\frac{1}{N}-T \label{4.0}%
\end{equation}
which corresponds to $I_{1}$.

May and Leonard note that the product $xyz $ (in our notation) has an
interesting asymptotic behaviour.

From (\ref{quad}) we find that
\begin{equation}
(XYZ)^{\prime}=-XYZ(1+\alpha+\beta)(X+Y+Z) \label{4.1}%
\end{equation}
so that
\[
\frac{(XYZ)^{\prime}}{XYZ}=-kN
\]
and in the case that $\alpha+\beta=2$ so that (\ref{sum1}) and (\ref{4.0})
apply this is
\[
\frac{(XYZ)^{\prime}}{XYZ}=3\frac{N^{\prime}}{N}%
\]
whence
\begin{equation}
\frac{XYZ}{(X+Y+Z)^{3}}=I_{2} \label{4.2}%
\end{equation}
which also may be written as
\begin{equation}
XYZ=I_{2}\left(  J_{1}+T\right)  ^{3}. \label{4.3}%
\end{equation}

The form (\ref{4.2}) indicates that this integral corresponds to the symmetry
$\partial_{T}$. The group theoretic origin of (\ref{4.2}) is easily seen. The
invariants of $\partial_{T}$ are $X$, $Y$ and $Z$. The requirement that
$f(X,YZ)$ be an integral of (\ref{quad}) leads to the associated Lagrange's
system
\[
\frac{\mbox{\rm d}X}{X^{\prime}}=\frac{\mbox{\rm d}Y}{Y^{\prime}}%
=\frac{\mbox{\rm d}Z}{Z^{\prime}}.
\]
We may use the theory of first-order differential equations \cite{Ince 27 a}
[p 45] to combine the elements in a specific fashion to give, when
(\ref{quad}) with $\alpha+\beta=2$ is taken into account,
\[
\frac{\mbox{\rm d}X}{X^{\prime}}=\frac{\mbox{\rm d}Y}{Y^{\prime}}%
=\frac{\mbox{\rm d}Z}{Z^{\prime}}=\frac{\mbox{\rm d}(XYZ)}{-3XYZ(X+Y+Z)}.
\]
Taking, say, the first with the fourth and using the equivalence of
$\mbox{\rm d}X/X^{\prime}$ to $\mbox{\rm d}T$ we have
\[
\frac{\mbox{\rm d}(XYZ)}{XYZ}=-3N\mbox{\rm d}T
\]
and (\ref{4.2}) follows when the composed equation, (\ref{sum1}), is used.

We may use (\ref{4.0}) and (\ref{4.3}) to eliminate $Y$ and $Z$ (say) from
system (\ref{quad}) with $\alpha+\beta=2$. To maintain a certain compactness
of notation we write $X+Y+Z$ as $N(T)$ and $XYZ$ as $m(T)$. We obtain
\begin{equation}
Z=\frac{m}{XY}\quad\mbox {\rm and}\quad Y^{2}+(X-N)Y+\frac{m}{X}=0 \label{4.4}%
\end{equation}
so that
\begin{equation}
Y=\mbox{$\frac{1}{2}$}\left\{  N-X\pm\sqrt{(N-X)^{2}-\frac{4m}{X}}\right\}.
\label{4.5}%
\end{equation}
The first-order differential equation satisfied by $X$ is found to be
\begin{equation}
X^{\prime}=-aX\mp\mbox{$\frac{1}{2}$}(\alpha-\beta)X\sqrt{(a-X)^{2}-\frac
{4b}{X}}. \label{4.6}%
\end{equation}

The single first-order differential equation, (\ref{4.6}), for $X (T) $
encapsulates the information already gleaned by our various analyses. The
contents of the root ensure our inability to find a solution to (\ref{4.6})
unless $\alpha=\beta$. When $\alpha=\beta=1 $, (\ref{4.6}) is always
integrable since then (\ref{4.6}) is simply a linear first-order equation.

It would be evident to a reader with only a modest acquaintance with symmetry analysis that the procedure of this Section is not unique.  We could look to replace the system (\ref {quad}) by either a single third-order differential equation or a second-order differential equation plus first-order differential equation which is a standard part of the method of reduction of order \cite {nucci96a, nucci01a}.

For the system (\ref {quad}) the former option is not feasible as leads to a very awkward algebraic equation.  The latter option is feasible and is found in the analysis of the Euler-Poinsot system \cite {Nucci 03 a}.  Indeed it is a logical consequence of the case for which two equations are replaced by a single second-order ordinary differential equation \cite {Nucci 05 a, Maharaj 06 a, Nucci 06 a}.  In the papers cited the symmetries of the second-ordered differential equations sought were Lie point symmetries.  In their study of the two-dimensional predator-prey system with malthusian growth Baumann and Freyberger \cite {Baumann 91 a} replace their two first-order differential equations with a polynomial second-order differential equation and then seek generalised symmetries with specific structure of the second-order equation.  These give generalised -- equally point since the two coalesce for systems of first-order equations -- symmetries of the original Lotka-Volterra system.  From the symmetries integrals and invariants follow easily.

We have not follow the procedure of Baumann and Freyberger in this paper since our investigation is, as the title of the paper proclaims, of analytic solutions and the range of parameters is already set by the singularity analysis.  Were our intentions otherwise, a symmetry analysis along the lines of those in the papers cited above would be appropriate.  The problem with (systems of) first-order equations is that the number of Lie point symmetries (equivalent, as noted above, to generalised symmetries) is infinite and so there is no finite algorithm for their determination.  The increase of order, which is an integral component of the method of reduction of order, makes it possible to implement a finite algorithm.  However, if like Baumann and Freyberger one introduces generalised symmetries at the higher order, the finite algorithm is lost.  One may as well substitute Ans\"atze of choice into the original system of first-order equations.

Our concern with the analytic behaviour of the system led us to use a rather restricted symmetry approach to the determination of appropriate symmetries.  If one removes the requirement of analycity, a wider investigation, even as general as that of Baumann and Freyberger, would be appropriate.

\section{Conclusion}

Our investigation of the model for competition among three species presented
by May and Leonard was motivated by the singularity and symmetry analyses
which are appropriate to integrable systems. We found that the critical values
of the parameters revealed in the analysis of the system using the methods of
dynamical systems were echoed in subsequent analyses from the viewpoint of the
singularity and symmetry approaches. That (1), \textit{\ae q} (\ref{quad}), is
a decomposed system when $\alpha+\beta=2$ made our analysis easier since the
composed equation, (\ref{sum1}), is trivially integrable. This provided one
invariant for system (\ref{quad}). The second invariant was suggested by the
analysis of May and Leonard who showed that $XYZ$ was expressible in terms of
an invariant for all values of $\alpha$ and $\beta$ provided that the
populations were small. When $\alpha=\beta=1$, the asymptotic invariant
becomes a global invariant. The existence of these two invariants,
corresponding to the two symmetries of invariance under time translation and
similarity transformation (more evident in (\ref{quad}) than (1)), enabled the
system of three autonomous first-order differential equations to be reduced to
a single nonautonomous first-order differential equation. The consequence of a
lack of further symmetry for general values of $\alpha$ and $\beta$ even with
the constraint $\alpha+\beta=2$ is quite evident in the form of the equation
with $\alpha+\beta=2$. The system (\ref{quad}) cannot exhibit chaos since the
autonomous invariant (\ref{4.2}) implies that the system can be reduced to an
autonomous system of order two and is thereby formally integrable. However,
this integrability is not in terms of an analytic function let alone in closed
form\footnote{We do not into a discussion of the meaning of integrability in
terms of functions which are not analytic. Although the formal requirement
that a solution be analytic is accepted, there are sufficient acceptable
exceptions for a certain laxness in practice.}. This is quite evident from the
form of (\ref{4.6}). The improvement in the integrability of (\ref{4.6}) as
the parameters become more closely aligned to the values which give favorable
results for the singularity analysis is clearly apparent. The simple removal of the imaginary part
of the resonances when $\alpha=\beta$ enables (\ref{4.6}) to be integrated
trivially in terms of an analytic function. In this respect the system
(\ref{quad}) is an excellent paradigm for the implications of the requirements
of the singularity analysis.

May and Leonard rightly indicate the marked change in the behaviour of the
system as the dimensionality is increased from two to three. The numerical
results which they present for $\alpha+\beta> 2 $ are very suggestive of the
behaviour in the solution which one would expect when the resonances are
complex. Indeed one would expect similar behaviour for $\alpha+\beta\leq2 $,
$\alpha\not =\beta$, but the change in sign of the real part of the exponent
means that there is damping of the oscillations rather than growth. Indeed, if
one considers the basis of the model, the values of the parameters $\alpha$
and $\beta$ should be such that $\alpha+\beta$ is likely to be an order of
magnitude less than two.

The community matrix of system (1)/(\ref{quad}) is circulant and a general
expression for the eigenvalues is available. When $\alpha=\beta$, the
community matrix becomes symmetric. Although May and Leonard make the point
that one could scarcely be interested at looking at the analysis of a system
of dimension greater than five, there may be some merit in the consideration
of systems of dimension greater than three. If all of the nondiagonal elements
are equal, which is a fairly drastic extension of the $\alpha=\beta$ case
discussed above, all submatrices containing the principal diagonal are
circulant and the possibility of the existence of composed systems leading to
invariants is real and integrability for competition among $n$ species is
conceivable. The constraint on the interaction coefficients to make the
community matrix circulant and so system (1) amenable to some analysis may be
regarded as severe. However, as we observed in the Introduction, among species
of similar habit as well as habitat the interaction coefficients are likely to
be less dominant than the self-specific effects. This is not a small step from
assuming equality.

\section*{Acknowledgements}

PGLL thanks Professor and Mrs Miritzis for their generous hospitality while
this work was undertaken and the University of KwaZulu-Natal for its
continuing support.

\end{document}